\documentclass[11pt]{article}
\usepackage[margin=1.25in]{geometry}
\usepackage{fancyhdr}
\usepackage{xspace}
\usepackage{hyperref}
\usepackage{color}
\usepackage{multirow}
\usepackage{authblk}
\usepackage{booktabs}
\usepackage{tabularx}
\usepackage{enumitem}
\usepackage{pdfpages}
\usepackage{graphicx}
\usepackage{subcaption}
\usepackage{float}
\usepackage{microtype}
\usepackage{xcolor}
\usepackage[
  shortcuts,%
  acronym,%
  nonumberlist,%
  nogroupskip,%
  nopostdot]{glossaries}

\usepackage[
  detect-none,%
  binary-units]{siunitx}[=v2]

\usepackage[
  sorting=none,%
  citestyle=numeric-comp,%
  bibstyle=ieee,%
  minbibnames=2,%
  maxbibnames=2,%
  maxcitenames=2,%
  mincitenames=1,%
  backend=biber,%
  isbn=false,%
  autolang=none,%
  bibencoding=auto]{biblatex}

\newenvironment{Abstract}{
    \begin{quotation}
        \begin{center}
            ABSTRACT
        \end{center}
        \bigskip
    \end{quotation}
}

\newcommand\snowmass{
    \begin{center}
        \rule[-0.2in]{\hsize}{0.01in}\\
        \rule{\hsize}{0.01in}\\
        \vskip 0.1in Submitted to the Proceedings of the US Community Study\\ 
        on the Future of Particle Physics (Snowmass 2021)\\ 
        \rule{\hsize}{0.01in}\\
        \rule[+0.2in]{\hsize}{0.01in}
    \end{center}
}

\def\Acknowledgements{
    \bigskip
    \bigskip
    \begin{center}
        \begin{large}
            \bf ACKNOWLEDGEMENTS 
        \end{large}
    \end{center}
}

\newcommand{\celeritas}{\emph{Celeritas}\xspace}

\newcommand{\acceleritas}{\emph{Acceleritas}\xspace}

\newcommand{\amd}{AMD\xspace}
\newcommand{\cuda}{CUDA\xspace}

\newcommand{\intel}{Intel\xspace}
\newcommand{\nvidia}{NVIDIA\xspace}
\newcommand{\oneapi}{OneAPI\xspace}

\newcommand{\Cpp}{C\texttt{++}\xspace}

\newcommand\registered{\textsuperscript{\textregistered}}

\newcommand{\ipl}[1]{(\textsl{\romannumeral #1})}

\DeclareSIUnit\flop{Flops}

\setacronymstyle{long-short}
\newacronym{adept}{AdePT}{Accelerated demonstrator of electromagnetic Particle Transport}
\newacronym{adios}{ADIOS}{Adaptable Input Output System}
\newacronym{ai}{AI}{artificial intelligence}
\newacronym{alcf}{ALCF}{Argonne Leadership Computing Facility}
\newacronym{anl}{ANL}{Argonne National Laboratory}
\newacronym{aosa}{AOSA}{Array of Structured Arrays}
\newacronym{api}{API}{Application Programming Interface}
\newacronym{ascr}{ASCR}{Advanced Scientific Computing Research}
\newacronym{atlas}{ATLAS}{A Toroidal LHC ApparatuS}

\newacronym{bp4}{BP4}{binary-pack V4}
\newacronym{bp5}{BP5}{binary-pack V5}
\newacronym{bnl}{BNL}{Brookhaven National Laboratory}
\newacronym{bsm}{BSM}{Beyond Standard Model}

\newacronym{cce}{CCE}{Center for Computational Excellence}
\newacronym{cern}{CERN}{Conseil Europ\'een pour la Recherche Nucl\'eaire}
\newacronym{ci}{CI}{Continuous Integration}
\newacronym{cms}{CMS}{Compact Muon Solenoid}
\newacronym{csg}{CSG}{Constructive Solid Geometry}
\newacronym[longplural={Central Processing Units}]{cpu}{CPU}{Central Processing Unit}

\newacronym{doe}{DOE}{U.S. Department of Energy}
\newacronym{dop}{DOP}{Data-Oriented Programming}
\newacronym{dune}{DUNE}{Deep Underground Neutrino Experiment}

\newacronym{ecp}{ECP}{Exascale Computing Project}
\newacronym{effis}{EFFIS}{Exascale Framework for High Fidelity coupled Simulations}
\newacronym{em}{EM}{electromagnetic}

\newacronym{fair}{FAIR}{Findable, Accessible, Interoperable, and Reusable}
\newacronym{fnal}{Fermilab}{Fermi National Accelerator Laboratory}
\newacronym[longplural={Figures of Merit}]{fom}{FOM}{Figure of Merit}

\newacronym{gdml}{GDML}{Geometry Description Markup Language}
\newacronym{gpu}{GPU}{Graphics Processing Unit}
\newacronym{gpfs}{GPFS}{General Parallel File System}

\newacronym{hep}{HEP}{High Energy Physics}
\newacronym{hip}{HIP}{Heterogeneous-Computing Interface for Portability}
\newacronym{hllhc}{HL-LHC}{High Luminosity Large Hadron Collider}
\newacronym{hpc}{HPC}{high performance computing}

\newacronym{io}{I/O}{input/output}

\newacronym{jlse}{JLSE}{Joint Laboratory for System Evaluation}
\newacronym{json}{JSON}{JavaScript Object Notation}

\newacronym{lanl}{LANL}{Los Alamos National Laboratory}
\newacronym{lbnl}{LBNL}{Lawrence Berkeley National Laboratory}
\newacronym{lcrc}{LCRC}{Laboratory Computing Resource Center}
\newacronym[longplural={Leadership Computing Facilities}]{lcf}{LCF}{Leadership Computing Facility}
\newacronym{llnl}{LLNL}{Lawrence Livermore National Laboratory}
\newacronym{lhc}{LHC}{Large Hadron Collider}

\newacronym{qa}{QA}{Quality Assurance}

\newacronym{mc}{MC}{Monte Carlo}
\newacronym{ml}{ML}{machine learning}
\newacronym{mpi}{MPI}{Message Passing Interface}

\newacronym{nersc}{NERSC}{National Energy Research Scientific Computing
    Center}
\newacronym{nsf}{NSF}{National Science Foundation}
\newacronym{nvme}{NVMe\registered}{Non-Volatile Memory Express\registered}
\newacronym{nvm}{NVM}{non-volatile memory}

\newacronym{olcf}{OLCF}{Oak Ridge Leadership Computing Facility}
\newacronym{openmp}{OpenMP}{Open Multi-Processing}
\newacronym{orange}{ORANGE}{Oak Ridge Adaptable Nested Geometry Engine}
\newacronym{ornl}{ORNL}{Oak Ridge National Laboratory}

\newacronym{pi}{PI}{Principal Investigator}
\newacronym{pii}{PII}{Personally Identifiable Information}
\newacronym{pps}{PPS}{Portable Parallelization Strategies}
\newacronym{prng}{PRNG}{pseudo-random number generator}
\newacronym{p5}{P5}{Particle Physics Project Prioritization Panel}

\newacronym{rapids}{RAPIDS2}{Resource and Application Productivity through
Computation, Information and Data Science}

\newacronym{sc}{SC}{Office of Science}
\newacronym{scidac}{SciDAC}{Scientific Discovery through Advanced Computing}
\newacronym{simd}{SIMD}{single instruction, multiple data}
\newacronym{simt}{SIMT}{single instruction, multiple thread}
\newacronym{sm}{SM}{Standard Model}
\newacronym{snl}{SNL}{Sandia National Laboratories}
\newacronym{sqe}{SQE}{Software Quality Engineering}
\newacronym{sst}{SST}{Sustainable Staging Transport}
\newacronym{surf}{SURF}{Sanford Underground Research Facility}

\newacronym{ui}{UI}{User Interfaces}
\newacronym{vecgeom}{VecGeom}{Vectorized Geometry}
\newacronym{vtkm}{VTK-m}{Visualization Toolkit-m}

\newacronym{wlcg}{WLCG}{Worldwide LHC Computing Grid}

\definecolor{CiteColor}{rgb}{0, 0, 0.55}
\definecolor{LinkColor}{rgb}{0.1, 0.1, 0.1}
\definecolor{URLColor}{rgb}{0.62745098, 0.1254902 , 0.94117647}
\definecolor{DraftColor}{RGB}{34, 139, 34}
\hypersetup{
  linkcolor=LinkColor,
  citecolor=CiteColor,
  urlcolor=URLColor,
  colorlinks=true
}

\newcommand\snowmassgroup{CompF2 -- Beam and Detector Simulation}
\newcommand\pubdate{\today}

\fancypagestyle{plain}{%
   \fancyhead[L]{
       \snowmassgroup
       \hfill
       \pubdate
    }
}

\title
{%
 \celeritas: GPU-accelerated particle transport for detector simulation in High
 Energy Physics experiments \footnote{This manuscript has been authored by
 UT-Battelle, LLC, under contract DE-AC05-00OR22725 with the US Department of
 Energy. The United States Government retains and the publisher, by accepting
 the article for publication, acknowledges that the United States Government
 retains a nonexclusive, paid-up, irrevocable, worldwide license to publish or
 reproduce the published form of this manuscript, or allow others to do so, for
 United States Government purposes. DOE will provide access to these results of
 federally sponsored research in accordance with the DOE Public Access Plan
  (\url{http://energy.gov/downloads/doe-public-access-plan}).}%
}
\date{}

\author[1]{S. C.~Tognini%
  \footnote{Corresponding author,
  \href{mailto:togninis@ornl.gov}{togninis@ornl.gov}}
}
\author[2]{P.~Canal}
\author[1]{T. M.~Evans}
\author[2]{G.~Lima}
\author[3]{A. L.~Lund}
\author[1]{S. R.~Johnson}
\author[2]{S. Y.~Jun}
\author[4]{V. R.~Pascuzzi}
\author[3]{P. K.~Romano}

\affil[1]{Oak Ridge National Laboratory, Oak Ridge, TN 37831, USA}
\affil[2]{Fermi National Accelerator Laboratory, Batavia, IL 60510, USA}
\affil[3]{Argonne National Laboratory, Lemont, IL 60439, USA}
\affil[4]{Brookhaven National Laboratory, Upton, NY 11973, USA}

\begin{document}
\maketitle

\begin{Abstract}
Within the next decade, experimental \ac{hep} will enter a new era of scientific
discovery through a set of targeted programs recommended by the \ac{p5},
including the upcoming High Luminosity \ac{lhc} \acs{hllhc} upgrade and the
\ac{dune}. These efforts in the Energy and Intensity Frontiers will require an
unprecedented amount of computational capacity on many fronts including \ac{mc}
detector simulation. In order to alleviate this impending computational
bottleneck, the \celeritas \ac{mc}  particle transport code is designed to
leverage the new generation of heterogeneous computer architectures, including
the exascale computing power of \ac{doe} \acp{lcf}, to model targeted \ac{hep}
detector problems at the full fidelity of Geant4. This paper presents the
planned roadmap for \celeritas, including its proposed code architecture,
physics capabilities, and strategies for integrating it with existing and future
experimental \ac{hep} computing workflows.
\end{Abstract}

\snowmass

\section{Introduction}

\ac{hep} is entering an exciting era for potential scientific discovery. A
targeted program, as recommended by the \ac{p5} report
\cite{ritz_building_2014}, addresses the main science drivers in all three
\ac{doe} \ac{hep} Frontiers: Energy, Intensity, and Cosmic. Two of these
flagship projects are the upcoming high luminosity upgrade of the \acl{lhc}
(\acs{hllhc}) and its four main detectors, and \ac{dune} at the \ac{surf} and
\ac{fnal}. These detectors will achieve a much higher granularity while also
being exposed to higher readout rates than previous experiments, leading to an
increase in data volume and complexity by orders of magnitude. As a consequence,
there are challenges in both online and offline computing infrastructures that
need to be overcome in order to store, process, and analyze this data.

Precision measurements and new physics discoveries rely heavily on comparisons
between recorded data and highly detailed \ac{mc} simulations. The required
quantity of simulated \ac{mc} data can be ten times greater than the
experimental data to reduce the influence of statistical effects and to study
the detector response over a very large phase space of new phenomena.
Additionally, the increased complexity, granularity, and readout rate of the
detectors require the most accurate---and thus most compute intensive---physics
models available. Therefore, these new facilities will require a commensurate
increase in computational capacity for the \ac{mc} detector simulations
necessary to extract new physics. However, projections of the computing capacity
available in the coming decade fall far short of the estimated capacity needed
to fully analyze the data from the \acs{hllhc}
\cite{the_atlas_collaboration_atlas_2020,cms-offline-computing-results}. The
contribution to this estimate from \ac{mc} full detector simulation is based on
the performance of the current state-of-the-art and \acs{lhc} baseline \ac{mc}
application Geant4 \cite{geant4,geant4_2006,geant4_status_2016}, a threaded
\acs{cpu}-only code whose performance has stagnated with the deceleration of
clock rates and core counts in conventional processors. Overcoming this
bottleneck by improving the performance of Geant4 or replacing components of its
simulation with lower-fidelity ``fast'' models, such as FastCaloSim
\cite{fastcalosim}, or \ac{ai}-based methods such as FastCaloGAN
\cite{fastcalogan}, is seen as a critical pathway to fulfilling the
computational requirements of the \acs{hllhc}
\cite{the_hep_software_foundation_roadmap_2019}.

Instead of relying on fast models to replace full-fidelity \ac{mc} transport, we
propose to break through the computational bottleneck with \celeritas
\cite{github-celeritas}, a new \ac{gpu}-optimized code to run full-fidelity
\ac{mc} simulations of \acs{hep} detectors. General-purpose compute accelerators
offer far higher performance per watt than \acp{cpu}.  \acp{gpu} are the most
common such devices and have become commodity hardware at the \ac{doe} \acp{lcf}
and other institutional-scale computing clusters. Of the top 10 performing
supercomputers in the current TOP500 list \cite{top500}, all are based on
accelerators and seven use \nvidia \acp{gpu}. However, adapting scientific codes
to run effectively on \ac{gpu} hardware is nontrivial and indeed has been the
goal of the multi-billion-dollar \ac{doe} \ac{ecp} \cite{ecp2019}. The
difficulty in adaptation results both from core algorithmic properties of the
physics and from implementation choices over the history of an existing
scientific code. The high sensitivity of \acp{gpu} to memory access patterns,
thread divergence, and device occupancy makes effective adaptation of \ac{mc}
physics algorithms especially challenging. Existing \ac{cpu} physics codes such
as Geant4 are impossible to port directly to vendor-independent \ac{gpu}
programming models due to common \Cpp language idioms such as polymorphic
inheritance and dynamic memory allocation. Therefore, instead of porting
existing codes, \celeritas is being developed from the outset.

A successful example of this approach is provided by the \ac{doe}'s \ac{ecp}.
During the course of the \ac{ecp} (2016--2023), 24 open science applications
have been ported and optimized for deployment on \ac{doe} \acp{lcf} with the
requirement that each meet specific performance metrics.  In all cases
\cite{evans_survey_2021}, this effort required substantial code redesign
involving both algorithms and data management in order to achieve performance on
the \ac{gpu}-based architectures that constitute the next generation of exascale
computing resources at the \acp{lcf}. For most applications, this resulted in
complete rewrites of the core solvers.

In this paper we present a roadmap for \celeritas. We describe the code
architecture features, the physics and geometry implementation, and our plan to
integrate \celeritas and \ac{doe} \acp{lcf} in current and future \ac{hep}
workflows. Our primary goal is to provide a tool for efficiently using all
available resources in \ac{hep} computing facilities by at least partially
offloading \ac{mc} production, which is one of the most intensive computing
tasks in an experiment, to the network of \acp{lcf}. Although it primarily aims
to reduce the computational demand of the \acs{hllhc}, we also envision
\celeritas being applied to the Intensity Frontier on \ac{dune}, maximizing the
use of advanced architectures that will form the backbone of \ac{hpc} over the
next decade.

\subsection{\ac{mc} simulation on \acsp{gpu}}

The first \ac{hep} \ac{mc} effort to exploit data-level parallelism, the
\emph{GeantV} project, targeted \ac{cpu} vector processing units as an untapped
source of computing power in deployed systems. While it demonstrated that
improving data and code locality could substantially speed up the simulation
software \cite{GeantV_Results_2020}, it also showed that simple vectorization is
insufficient to achieve the concurrency needed for transformative performance
gains. Modern \acp{gpu} offer significantly more opportunities for parallelism
than \ac{cpu} vector processing units and have a more flexible programming
paradigm that will allow performance improvements well beyond the results seen
in \emph{GeantV}.

A recently developed \ac{gpu} code in the \ac{hep} \ac{mc} space, Opticks,
leverages the \nvidia ray tracing library OptiX to simulate photon interactions
in liquid argon detectors \cite{blyth_opticks_2019}. The extremely high (tens of
millions) number of photons in flight created by a single detector interaction
and relatively simple physics makes this problem an ideal candidate for \ac{gpu}
acceleration, and indeed the code provides a speedup in the factor of hundreds
compared to transporting the emitted photons through Geant4. While Opticks was
designed to take advantage of unique architectural features available on
\nvidia, the problem being solved there is highly specialized.  The requirements
for tracking multiple particle types through complex geometries in the presence
of external magnetic fields is not well suited to the ray-tracing approach used
in Opticks.  Supporting Energy and Intensity Frontier experiments in \ac{hep}
requires a broader set of physics and capabilities.

Shortly after the start of preliminary work on \celeritas, \acs{cern} launched a
new effort, \acs{adept}, to evaluate the performance potential of \ac{em} shower
simulation on \acp{gpu} \cite{andrei_gheata_adept_2020}. Initial
test problem results \cite{andrei_gheata_adept_2021,hahnfield_2021} have
demonstrated performance parity between a 24-core Geant4 simulation and an
\acs{adept} simulation on a single consumer-level \ac{gpu}. Both \acs{adept} and \celeritas leverage
the \acs{vecgeom} geometry navigation library, a product of the \emph{GeantV}
project designed for \ac{cpu} vectorization that has since been extended to
\ac{gpu} multithreading using \cuda \cite{apostolakis_towards_2015}.

The proposed approach in \celeritas differs substantially from \emph{GeantV} and
\acs{adept}. Whereas \acs{adept} is seeking to implement \ac{gpu} capability in
Geant4 through minimal algorithmic and data refactoring, the \celeritas code
architecture is designed from the outset to support algorithms and data layouts
that are optimized for the unique requirements of \ac{gpu} architectures. The
\ac{mc} particle transport method is characterized by high thread divergence,
random data access patterns, high code complexity to sample physics
interactions, and relatively low arithmetic intensity.  Achieving high kernel
occupancy to manage memory latency on \acp{gpu} requires strategies that yield
local memory collocation and optimal kernel register usage
\cite{hamilton_continuous-energy_2019}.  These features are difficult or
impossible to achieve simply by modifying code designed for cache-based
architectures.

Outside the realm of \ac{hep}, recent work in the \emph{ExaSMR: Coupled Monte
Carlo Neutronics and Fluid Flow Simulation of Small Modular Reactors} project
within \ac{ecp} has demonstrated equivalence of $160$ \ac{cpu} cores per
\ac{gpu} on
Summit\footnote {
    \url{https://www.olcf.ornl.gov/olcf-resources/compute-systems/summit/}
} for the Shift \ac{mc} transport code on full-featured, three-dimensional
reactor models \cite{hamilton_continuous-energy_2019}. There are important
differences between this work and the necessary capabilities required for
particle physics detector modeling. For instance, \emph{ExaSMR} applications are
not characterized by large showers of secondary particles, and because the
particles are neutral, there are no \ac{em} field interactions. Despite the more
complex needs of \ac{hep} \ac{mc} simulations, current preliminary results
presented on this paper show a promising outcome for this project, with a fully
fledged \celeritas potentially reaching equivalent \ac{gpu}-to-\ac{cpu}
performance.

\section{\celeritas code}

\subsection{Code Architecture}

\begin{figure}
  \centering
  \includegraphics[scale=.6]{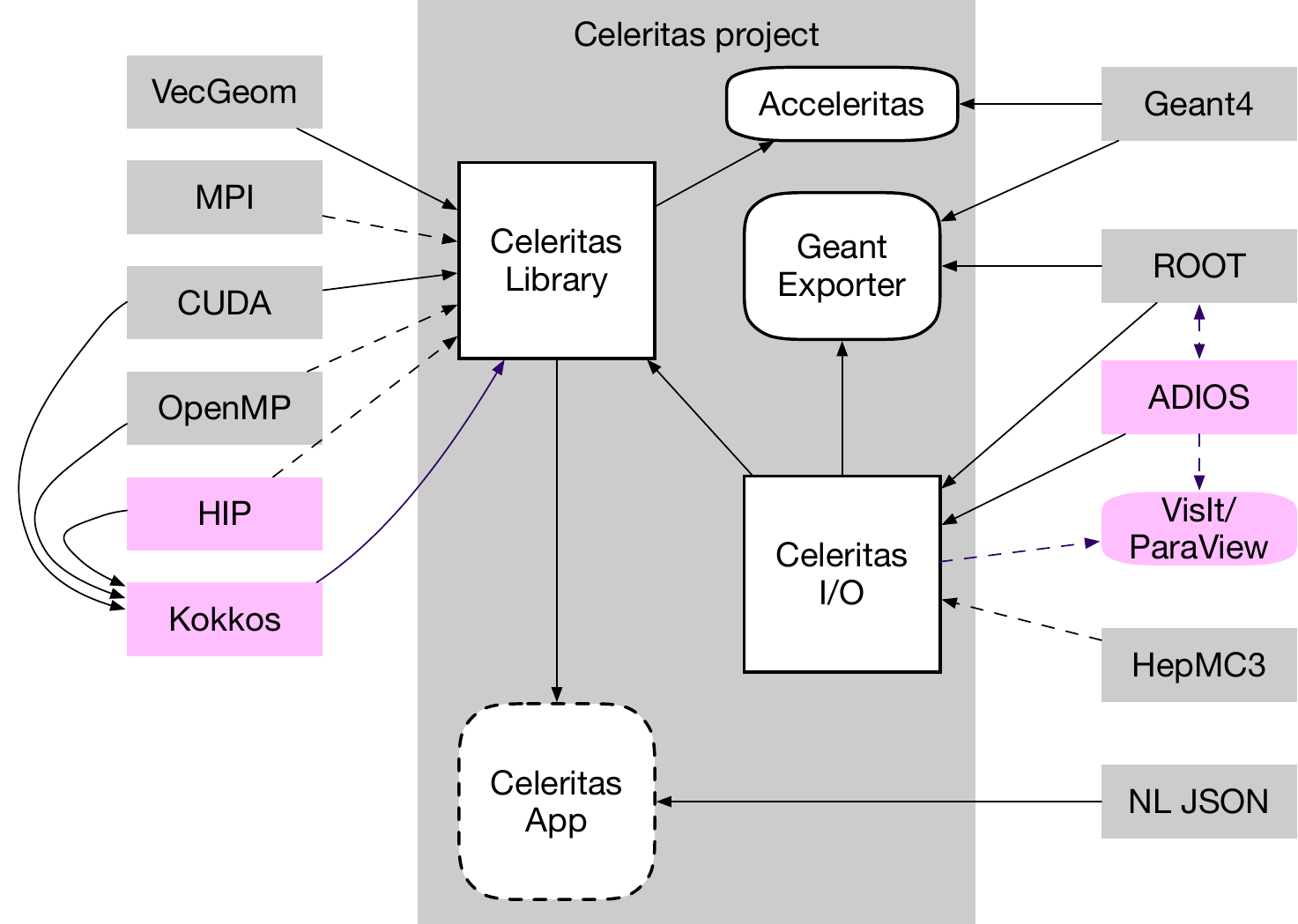}
  \caption{\celeritas code base (white) and its existing (gray) and proposed
    (magenta) third-party dependencies, both required (solid lines) and optional
    (dashed lines).}
  \label{fig:celeritas-code-base}
\end{figure}

A detailed description of the \celeritas code architecture is given in
\textcite{johnson_2021}. The code base (Fig.~\ref{fig:celeritas-code-base})
relies on external dependencies for key capabilities that are discussed in the
following sections.

\celeritas supports intra-node concurrency on multi-core architectures through
\acs{openmp} and on \nvidia \acp{gpu} using \cuda, which we plan to supplement
with a programming model for vendor-independent portability.  Internode
parallelism is currently implemented using the \ac{mpi} communication library
through a domain replication model in which particle events are decomposed
across \ac{mpi} ranks.

Like other \ac{gpu}-enabled \ac{mc} transport codes such as Shift
\cite{pandya_implementation_2016,hamilton_multigroup_2018,
hamilton_continuous-energy_2019,hamilton_domain_2022}, the low-level component
code used by transport kernels is designed so that each particle track
corresponds to a single thread. There is no cooperation between individual
threads, facilitating the dual host/device annotation of most of \celeritas.
\celeritas also uses a modular programming approach based on composition rather
than inheritance in order to accommodate device-based architectures, which have
poor support for runtime polymorphism.

The \celeritas programming model uses the \ac{dop} paradigm \cite{dop_2022} to
facilitate platform portability, improve memory access patterns, and accelerate
development. \ac{dop} separates execution code from data, and as part of this
model \celeritas carefully partitions immutable, shared ``parameter'' data from
dynamic thread-local ``state'' data. Object-oriented design patterns encapsulate
the data storage implementation, temporarily combining parameter and state data
into ``view'' classes.  Higher-level classes use composition to combine the data
from the multiple entities that comprise a particle track's complete state.

In the first 1.5 years of \celeritas' development, nine \ac{gpu}-compatible
physics models (Table~\ref{tab:em-physics}) have been implemented. This shows
\ac{dop} to be highly effective for development on heterogeneous architectures
that have independent \emph{memory spaces} between which data must be
transferred. One challenge faced by \ac{mc} physics application codes is the
ubiquity of complicated \emph{heterogeneous} data structures needed for
tabulated physics and particle data, as opposed to the simpler
\emph{homogeneous} data layouts required by deterministic numerical solvers. A
novel programming model in \celeritas enables the composition of new, deep data
types (e.g., material properties) that are required by geometric and physics
operations during the transport loop without fragmenting the underlying data
layout on device.

One requirement for transporting particles in \ac{em} showers is the efficient
allocation and construction of secondary particles during a physics interaction.
On \acp{gpu}, managing dynamic allocations efficiently is a significant
challenge.  To enable runtime dynamic allocation of secondary particles, we have
developed a function-like stack allocator that accesses a large on-device
allocated array with a fixed capacity and uses an atomic addition to
unambiguously reserve one or more items in the array. The final aspect of
\ac{gpu}-based secondary allocation is how to gracefully handle an out-of-memory
condition without crashing the simulation \emph{or} invalidating its
reproducibility. A novel algorithm in \celeritas guarantees robustness when
allocating secondaries, which we will extend to guarantee complete
reproducibility of \ac{hep} workflow results.

The \celeritas code architecture just summarized is designed to enable \ipl{1}
performance portability, \ipl{2} implementation of new physics models and
processes, \ipl{3} optimal geometric tracking, \ipl{4} optimization of the
particle transport algorithm in the presence of external \ac{em} fields, and
\ipl{5} addition of scoring and \ac{ui} necessary to meet \acs{hep} detector
simulation requirements.

\subsubsection{Platform portability}

The next generation platforms at the \ac{olcf}, \ac{alcf}, and \ac{nersc} will
each feature different \ac{gpu}-based architectures as shown in
Table~\ref{tab:lcf-arch}, so \celeritas cannot rely on a single proprietary
programming model. The code currently uses a limited set of macros and
automatically generated kernels to support \cuda, \acs{hip}, and \acs{openmp}.
Its highly modular data management design and function-like objects used to
launch kernels will allow for straightforward adaptation to other programming
models as needed.
\begin{table}[h]
  \caption{Exascale architectures and native programming models at \acs{doe}
  \acsp{lcf}.}
  \label{tab:lcf-arch}
  \centering%
  \begin{tabular}{lllll}\toprule
    Center & Machine & Integrator & \acsp{gpu} & Native Programming
    Model\\\midrule
    \acs{olcf} & Frontier & HPE & \amd & \acs{hip}\textsuperscript{a}\\
    \acs{alcf} & Aurora & \intel & \intel &
    \oneapi/DP\Cpp/SYCL\textsuperscript{b}\\
    \acs{nersc} & Perlmutter & HPE & \nvidia &
    \cuda\hspace{-.4em}\textsuperscript{c}\\
    \bottomrule
    \multicolumn{5}{l}{\footnotesize
      \textsuperscript{a}\url{https://github.com/ROCm-Developer-Tools/HIP} }\\
    \multicolumn{5}{l}{\footnotesize
      \textsuperscript{b}\url{https://software.intel.com} }\\
    \multicolumn{5}{l}{\footnotesize
      \textsuperscript{c}\url{https://docs.nvidia.com/cuda/cuda-runtime-api/index.html}
      }\\
  \end{tabular}
\end{table}

\textcite{evans_survey_2021} surveyed the various \ac{gpu} programming models
employed by applications within the \ac{ecp} and found Kokkos
\cite{CarterEdwards20143202}, \acs{hip}, and \acs{openmp} to be the most
commonly employed programming models.  Each of these models has pros and cons:
some models are not yet supported on all \ac{gpu} architectures, and experience
in \ac{ecp} has shown that performance can vary dramatically depending on the
maturity of both the software stack and the underlying hardware. This is also a
time of significant change in the \Cpp language itself, particularly with
respect to concurrency support via Standard Library algorithms. Combined with
the increasing adoption of LLVM for \Cpp compiler development, we anticipate
that more vendors will provide \ac{gpu}-based concurrency \acp{api} through \Cpp
Standard Library constructs within the next \numrange[range-phrase={ to
}]{5}{10} years, providing yet another possible means of achieving platform
portability.

The \celeritas team will evaluate and subsequently choose a programming model to
provide portable execution across all major \ac{gpu} vendors as well as
traditional multicore \acp{cpu}. This evaluation will be based on achievable
performance across multiple \ac{gpu} architectures, ease of integration into
\celeritas workflows, sustainability, availability of multiple implementations,
and ability to perform platform-specific tuning. We also plan to engage the
\ac{hep} \ac{cce} \ac{pps} working group to ensure our strategy is in line with
other efforts in the \ac{hep} community. At the present time, our nominal
performance portability plan for \celeritas is to utilize one of the \Cpp-based
programming models (Kokkos, SYCL, or \Cpp standard library execution policies);
however, we recognize that these models are rapidly changing as is compiler
support, and thus a proper evaluation is necessary before making a decision.

\subsection{Physics}

The key physics component in \celeritas is a \emph{process}, which defines an
observed physical phenomenon such as the photoelectric effect or bremsstrahlung.
Each process is implemented as one or more models that each mathematically
describe or approximate the process in a given energy regime.

The initial implementation in \celeritas targets \ac{em} physics between
\SI{100}{eV} and \SI{100}{TeV} for photons, electrons, and positrons. This
minimal set of capabilities, with physical processes and associated numerical
models itemized in Table~\ref{tab:em-physics}, is necessary to generate
realistic simulations of \ac{em} showers and demonstrate key characteristics of
a full-featured transport loop.
\begin{table}[h]
  \caption{Current status of \celeritas \acs{em} physics. The initial
  implementation ($\gamma$ and $e^\pm$) is almost complete, and muon physics is
  in its initial stage. Particle symbols are defined in
  \textcite{tanabashi_review_2018}.}
  \label{tab:em-physics}
  \centering
  \begin{tabular}{clll}\toprule Particle & Process & Model(s) & Status\\
    \midrule
    \multirow{4}{*}{$\gamma$}
    & photon conversion & Bethe--Heitler & implemented\\
    & Compton scattering & Klein--Nishina & verified\\
    & photoelectric effect & Livermore & implemented\\
    & Rayleigh scattering & Livermore & implemented\\
    \midrule
    \multirow{4}{*}{$e^\pm$}
    & ionization & M\o{}ller--Bhabha & implemented\\
    & bremsstrahlung & Seltzer--Berger, relativistic & implemented\\
    & pair annihilation & EPlusGG & implemented\\
    & multiple scattering & Urban, WentzelVI & in progress\\
    \midrule
    $\mu^\pm$ & muon bremsstrahlung & Muon Bremsstrahlung & implemented\\
    \bottomrule
  \end{tabular}
\end{table}
These already implemented characteristics include \ipl{1} material-dependent
physical properties, \ipl{2} continuous slowing down in matter for charged
particles, \ipl{3} selecting discrete interactions among competing processes,
\ipl{4} scattering or absorbing particles during an interaction, \ipl{5}
emitting secondary particles, and \ipl{6} applying energy cutoffs to cull
low-energy photons and electrons.

The physics implementation in \celeritas focuses on maximizing work done in
parallel. For example, all particle types use tabulated discrete interaction
cross sections calculated simultaneously in a single kernel. The primary
deviation from this rule is that each model of a discrete process launches an
independent kernel that applies only to tracks undergoing an interaction with
that process. This set of kernel launches is performed polymorphically from
\ac{cpu} host code, allowing arbitrary noninvasive extensions to \celeritas
physics.

In order to meet the detector simulation requirements for \acs{hep} experiments,
\celeritas physics will be expanded from its initial \ac{em} prototype to a full
set of particles with decay and hadronic physics.  A complete list of the
required physics processes and particles is shown in
Table~\ref{tab:proposed-physics}, where only processes are explicit as model
separation will be determined based on performance and/or code maintainability.
\begin{table}[h]
  \caption{Proposed physics development in \celeritas. Model definitions are
  ommited as these will be determined based on code performance and/or
  maintainability.}
  \label{tab:proposed-physics}
  \centering
  \begin{tabular}{llll}
    \toprule
    Physics & Process & Particle(s)\\
    \midrule
    \multirow{10}{*}{\acs{em}} & photon conversion & $\gamma$\\
    & pair annihilation & $e^\pm$\\
    & photoelectric effect& $\gamma$\\
    & ionization & charged leptons, hadrons, and ions\\
    & bremsstrahlung & charged leptons and hadrons\\
    & Rayleigh scattering & $\gamma$\\
    & Compton scattering & $\gamma$\\
    & Coulomb scattering & charged leptons, hadrons\\
    & multiple scattering & charged leptons, hadrons\\
    & continuous energy loss & charged leptons, hadrons, and ions\\
    \midrule
    \multirow{3}{*}{Decay}
    & two body decay & $\mu^\pm$, $\tau^\pm$, hadrons\\
    & three body decay & $\mu^\pm$, $\tau^\pm$, hadrons\\
    & n-body decay & $\mu^\pm$, $\tau^\pm$, hadrons\\
    \midrule
    \multirow{6}{*}{Hadronic}
    & photon-nucleus & $\gamma$ \\
    & lepton-nucleus & leptons \\
    & nucleon-nucleon & $p$, $n$\\
    & hadron-nucleon & hadrons\\
    & hadron-nucleus & hadrons\\
    & nucleus-nucleus & hadrons\\
    \bottomrule
  \end{tabular}
\end{table}

Geant4 manages cross sections by a combination of importing experimental values
from external databases and programmatic implementations of theoretical models.
Most of the data used in the simulation is pre-tabulated during initialization,
with some models calculating elemental cross sections at runtime. Unlike Geant4,
\celeritas currently does not have a fully-fledged system to load and
pre-tabulate cross sections from external databases. Therefore, cross section
tables are imported from Geant4 via an external application---the Geant4
exporter in Fig.~\ref{fig:celeritas-code-base}. This application loads the
problem geometry via a \acs{gdml} file, initializes Geant4, and stores all the
pre-calculated cross sections for all the available physics processes and models
into a ROOT file. This file is then used by a \celeritas application to load the
data to host and device at initialization time.

\subsection{Geometry and \ac{em} fields}

Accurate simulation of \ac{hep} detector output requires a highly detailed model
of the detector apparatus and nearby components. \ac{mc} transport requires that
the model geometry be well-defined at every point in space, requiring a model
that is ``watertight,'' or heuristics for recovering from inconsistencies (e.g.,
overlapping or missing regions), or both. Since particles undergo collisions and
charged particles constantly change direction as they move through the magnetic
fields in the detector, traditional straight-line ray tracing is necessary but
not sufficient to correctly navigate the geometry.

\begin{figure}
  \centering%
  \begin{subfigure}{3in}%
    \centering%
    \includegraphics[height=2in]{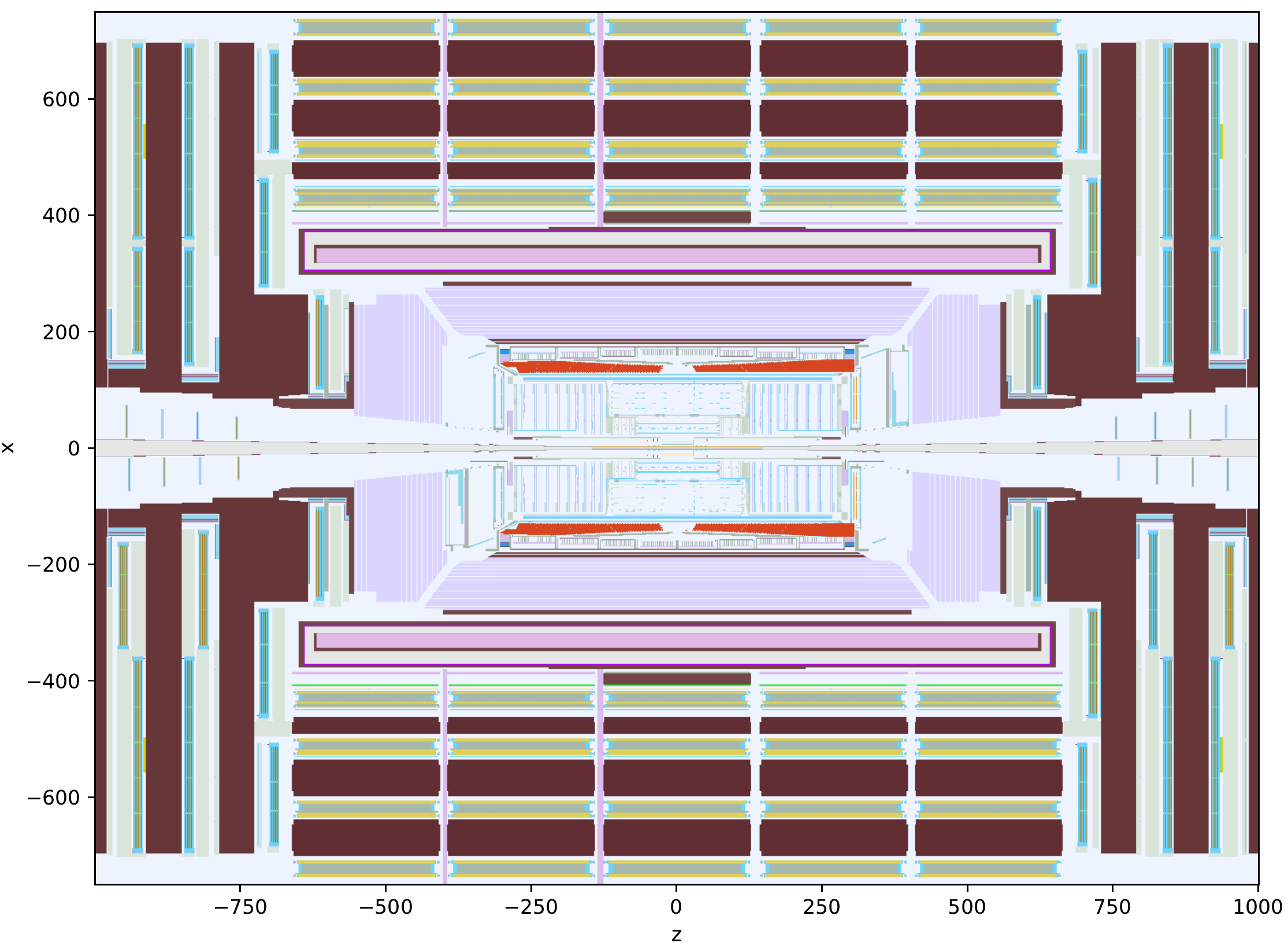}%
    \caption{$y=0$ [cm]}%
  \end{subfigure}%
  \begin{subfigure}{3in}%
    \centering%
    \includegraphics[height=2in]{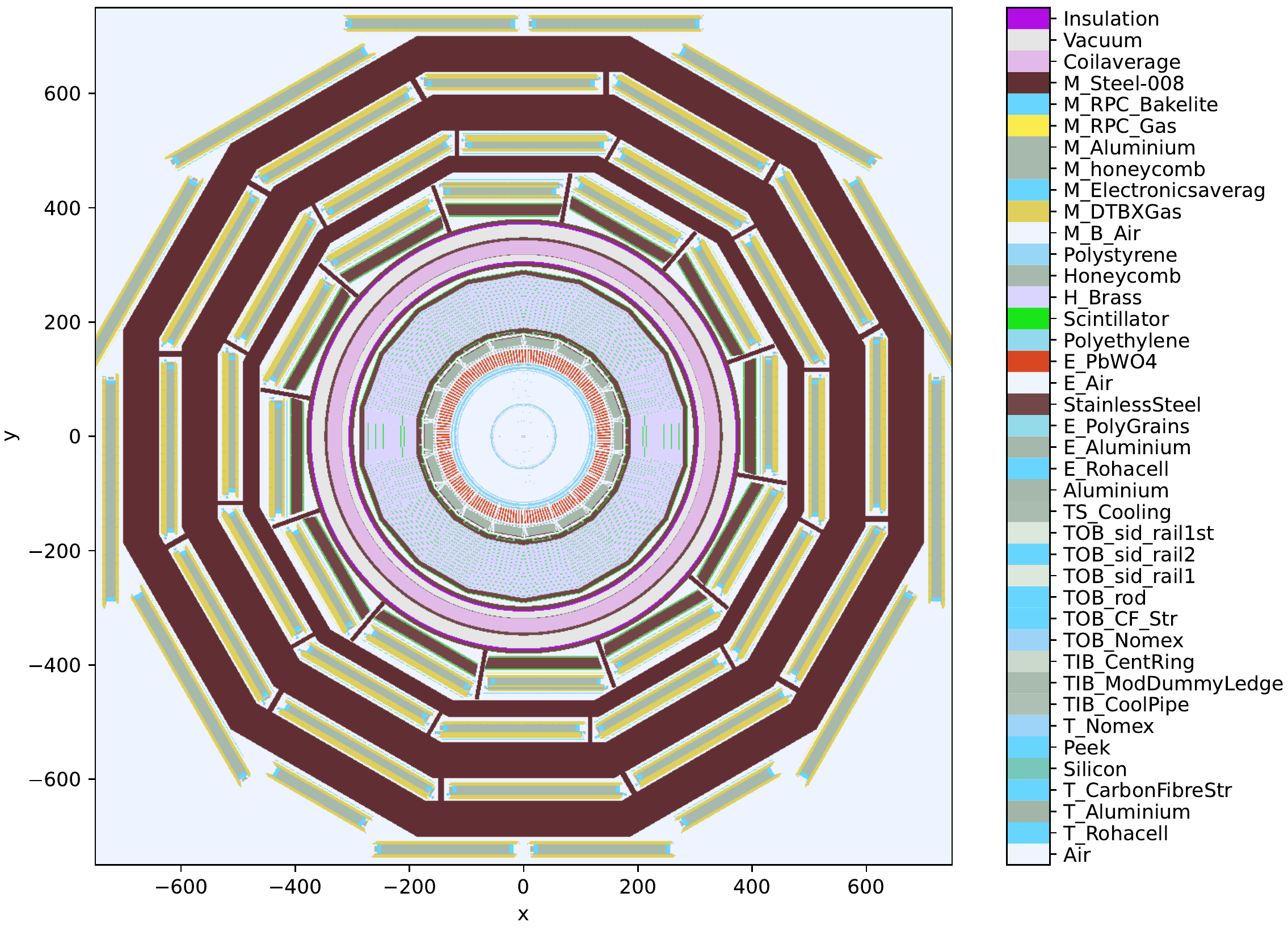}%
    \caption{$z=0$ [cm]}%
  \end{subfigure}
  \caption{On-device \acs{vecgeom} ray traces of the \acs{cms} generated with
the \celeritas ray trace demonstration app.}
\label{fig:vecgeom-cms}
\end{figure}

The \acs{vecgeom} library supports navigation through Geant4-defined detector
geometries in \ac{cpu}-only code and on \cuda-enabled devices
\cite{apostolakis_towards_2015} and is the initial and primary geometry package
in \celeritas. Figure~\ref{fig:vecgeom-cms} is a representation of the \ac{cms}
geometry traced in parallel on \ac{gpu} using the \celeritas particle tracking
interface to \acs{vecgeom}.

A new, alternate geometry in \celeritas provides a test bed for experimenting
with platform-portable navigation that uses fundamentally different algorithms
from \acs{vecgeom}. This implementation, \ac{orange}, is an initial \ac{gpu}
port of the new modernized geometry used by the SCALE nuclear engineering code
suite to model complex multi-level nuclear reactor and neutral particle
shielding problems~\cite{scale}. The \celeritas port uses the Collection
paradigm to store quadric surface representations and define cell volumes as
\ac{csg} combinations of those surfaces. At present the \ac{gpu} prototype
implementation supports only a single geometry level, but the extension of this
implementation to the full capabilities available on \ac{cpu} will be relatively
straightforward. We are actively collaborating with the \acs{vecgeom} group at
\acs{cern} to research how \ac{orange} and its methodology could power the next
generation of Geant4 tracking on \ac{gpu}.

\celeritas supports on-device propagation of particles through arbitrary
magnetic fields. Using a \Cpp-based template system for high-performance,
platform-independent extensibility, it allows different integration algorithms
for different user-defined fields.

\subsection{User-interface and \ac{io}}

The key challenge in providing user interfaces is developing a framework that
supports integration into existing experimental \ac{mc} simulation sequences
that provides the minimum required code barrier for incorporation and preserves
performance.  Existing experimental frameworks are built on the Geant4 toolkit,
which provides user actions that allow users to control the program and data
flow at the level of run, event, track, and step. In Geant4 scoring is done
using dedicated stepping actions in which information from the sensitive
detector volumes is accessible through callback semantics into these parts of
the simulation.  This approach provides great user flexibility at the cost of
higher computational overhead and increased system complexity. Furthermore,
callback functions will not work in accelerator code because it is not possible
to call host functions in the middle of device kernel execution.

\celeritas will not operate as a toolkit as Geant4 does, since this would leave
many implementation decisions to the end-user, hindering performance. Therefore,
to address the purely technical challenges of supporting experimental workflows,
we will implement an \ac{api} through which clients can specify geometric
regions for scoring and \ac{mc} particle data. Using this \ac{api} the desired
scoring data can be processed on the host at runtime, and the necessary data
fields for tallies can then be configured for execution in kernel code on the
device.  The most efficient interface would fully occupy the device by executing
many events concurrently.  However, this is not the way that experimental
workflows are currently configured, in which events are executed independently
on each thread.

As \ac{hpc} evolved to make use of heterogeneous architectures, \ac{io} became
one of the critical performance bottlenecks of many \ac{hpc} applications and is
a main concern for \celeritas. \ac{hep} detector simulations produce large
volumes of data, with many millions of particle tracks and detector scoring
regions having to be recorded. Thus, the need to optimize the data movement
between host and device and manage parallel \ac{io} requests is paramount.
Furthermore, to be compatible with \ac{hep} experimental workflows, \celeritas
needs to be integrated with ROOT. To address these challenges, we are
collaborating with \ac{doe}'s \ac{rapids} team to make \ac{adios} the internal
\ac{io} \ac{api} of \celeritas, as it is highly optimized for heterogeneous
architectures. The ultimate goal is to find optimal strategies to mitigate
\ac{io} performance issues and integrate \ac{adios} with ROOT for full
interoperability with \ac{hep} workflows. This collaboration with \ac{rapids}
will also explore data visualization tools and event filtering, allowing users
to visualize, validate, and debug the generated data before launching production
campaigns.

\section{\celeritas performance}

Given the range of detector geometry complexities, primary event types, and
amount of output, there is no general case that provides a simple performance
number such as a simulation rate in events per second. Therefore, our ultimate
performance metric will be based on a set of real-world \acs{hep} detector
workflow use cases. \celeritas must run the targeted models at the same fidelity
as the current state of the art but in much less time. In the next sections,  we
will present the two \acp{fom} that will be used to measure success, along with
a few preliminary performance results comparing \celeritas with current
state-of-the-art \ac{mc} codes.

\subsection{Performance metrics}

On \ac{lcf} hardware for the next decade, \acp{gpu} will provide the bulk of the
compute capability, so one critical performance metric is the ratio of the
runtime of \celeritas using \acp{gpu} to using only the \acp{cpu} of a given
machine. This \ac{fom} represents the ability of \celeritas to effectively use
the hardware that is available and must be sufficiently high to justify running
on \ac{lcf} resources. We will target a factor of $160\times$ for a single
\ac{gpu} to a single \ac{cpu}, which is the relative \ac{gpu}/\ac{cpu}
performance of the Shift \ac{mc} transport code
\cite{hamilton_continuous-energy_2019}.

A second performance metric is critical to the programmatic viability of
\celeritas in the broader \ac{hep} community: the work done for the same amount
of cost in power consumption and hardware, using Geant4 on \ac{cpu} as a
baseline. This \ac{fom} is the motivation for \ac{hep} workflows to adapt to
using \celeritas for \ac{mc} transport. The Geant4 team supposes that a factor
of two speedup resulting from ``adiabatic improvements'' to their code is not
outside the realm of possibility \cite{marc_verderi_geant4_2021}, so a $2
\times$ cost improvement of \celeritas runtime over Geant4 runtime is our second
target. Assuming that electricity consumption (and waste heat disposal) are the
primary constraints for independent \ac{hep} computing centers, this factor
should be evaluated by comparing the performance of Geant4 on \acp{cpu} with a
comparable power requirement as \celeritas on \acp{gpu}. At the present time,
for example, that comparison might be for an AMD 3rd Gen EPYC Processor power
(\SI{280}{\watt} for 64 cores) to a PCIe \nvidia A100 (\SI{250}{\watt} for 108
symmetric multiprocessors).

Meeting these \acp{fom} with Geant4-equivalent physics capabilities and
providing a solid user interface and \ac{io} integration to \ac{hep} workflows
will be the ultimate confirmation that \celeritas is a viable option for
\ac{hep} experiments. At this point \celeritas will be ideal for execution on
\ac{doe} \acp{lcf} and will be sufficiently fast on its own merits to motivate
independent adoption on capacity systems.

\subsection{Preliminary results}

The \celeritas team will develop a series of internal test problems to validate
every aspect of the code, including physics, geometry, \ac{em} fields, and
\ac{io}. Here, we present a small set of preliminary results comparing
\celeritas against Geant4 using our most recently implemented high-level
verification problem. This test evaluates the current code as a whole, including
features such as multiple materials, secondary production, and energy cutoffs
with the full array of currently implemented physics processes for $\gamma$,
$e^-$, and $e^+$ (Table~\ref{tab:em-physics}). The test geometry is an idealized
coarse representation of the main components of the \ac{cms} detector,
comprising six cylindrical shells centered about the $z$ axis up to
\SI{375}{\centi\meter}. Each cylinder and shell is composed of a single element:
vacuum, Si, Pb, C, Ti, and Fe.

The first preliminary results are on a compute node that uses the same \acp{gpu}
as \ac{olcf}'s Summit: it has Intel Xeon Gold 5218 (Cascade Lake) \acp{cpu}
running at 2.3GHz (2 sockets of 16 cores with 2 hardware threads per core, with
\SI{188}{\giga\byte} total memory), and \nvidia V100 \acp{gpu} (each with 80
symmetric multiprocessors, 64 CUDA cores per multiprocessor, and
\SI{16}{\giga\byte} memory). The simulation input is an isotropic
\SI{1}{\giga\electronvolt} photon point source at the origin, composed of 100
events each with \num{1000} photon primaries.

\begin{figure}
  \centering%
  \begin{subfigure}{\textwidth/2}%
    \includegraphics[height=2.6in]{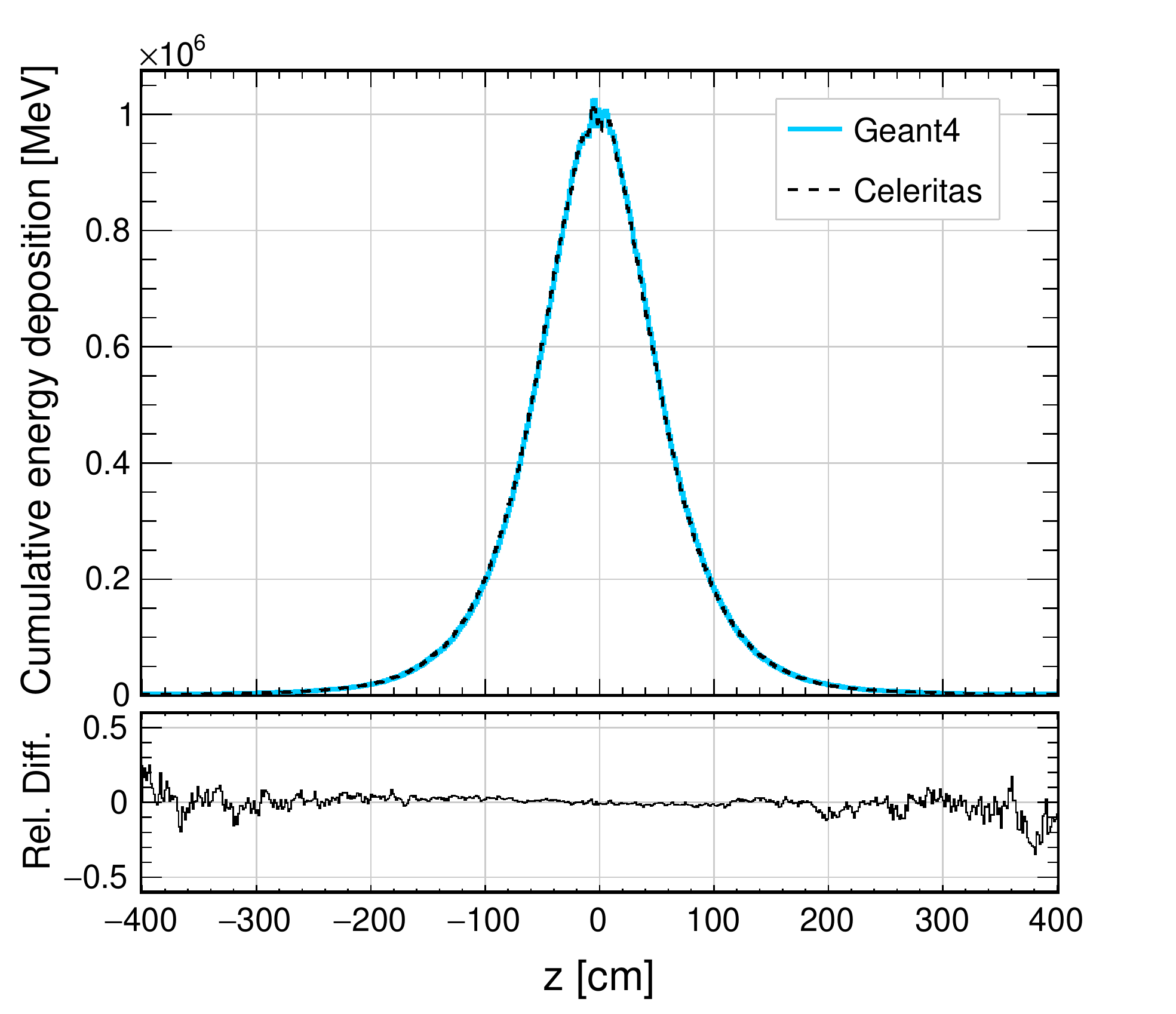}
    \caption{Cumulative energy deposition over the $z$ axis.}%
    \label{fig:physics-results:edep}
  \end{subfigure}%
  \begin{subfigure}{\textwidth/2}%
    \includegraphics[height=2.6in]{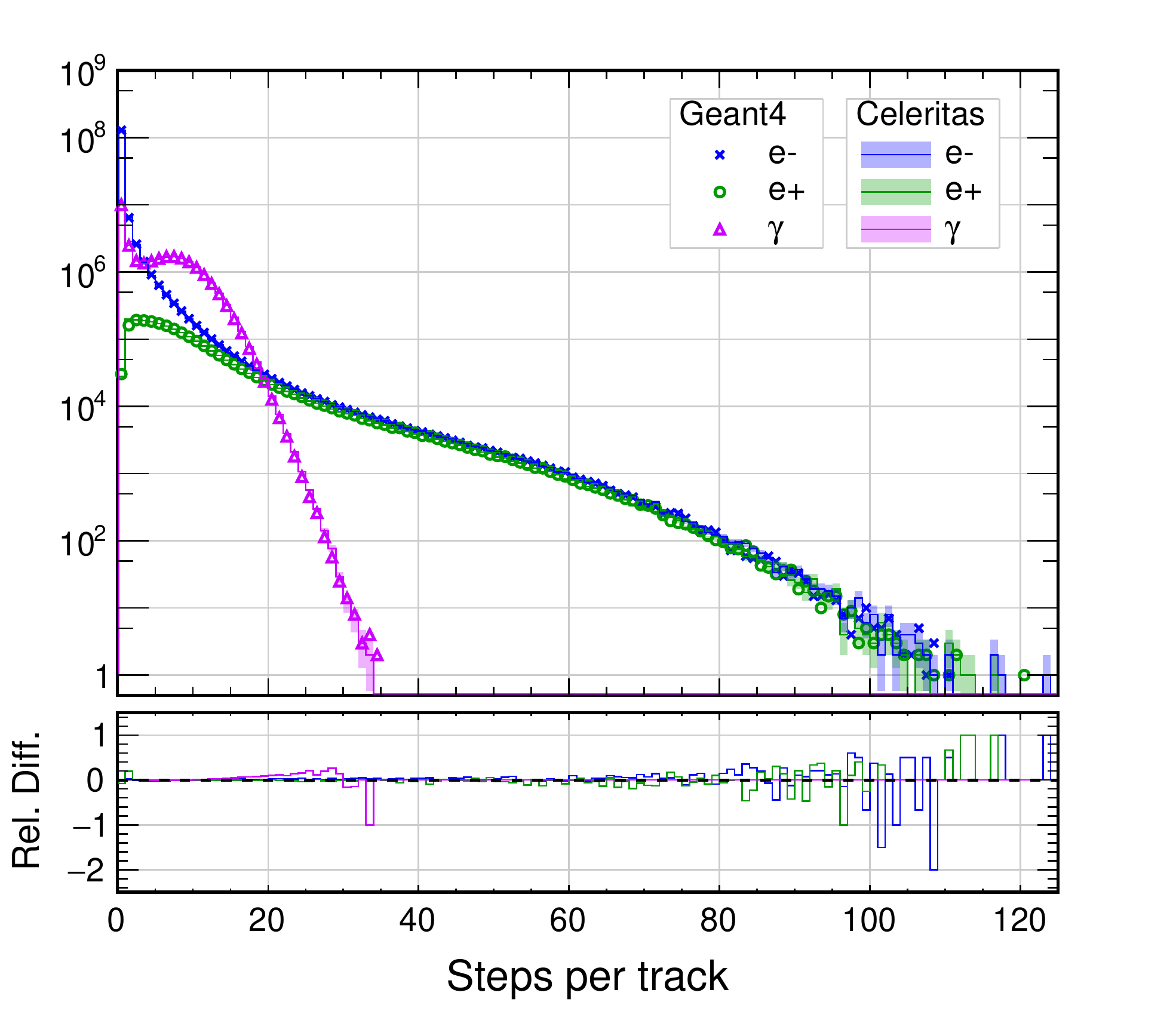}%
    \caption{Steps per track, separated by particle type.}%
    \label{fig:physics-results:steps}
  \end{subfigure}
  \caption{Comparison between Geant4 and \celeritas simulation runs. Results are
  in agreement within the statistical errors.}
  \label{fig:physics-results}
\end{figure}

\begin{figure}
    \centering
    \includegraphics{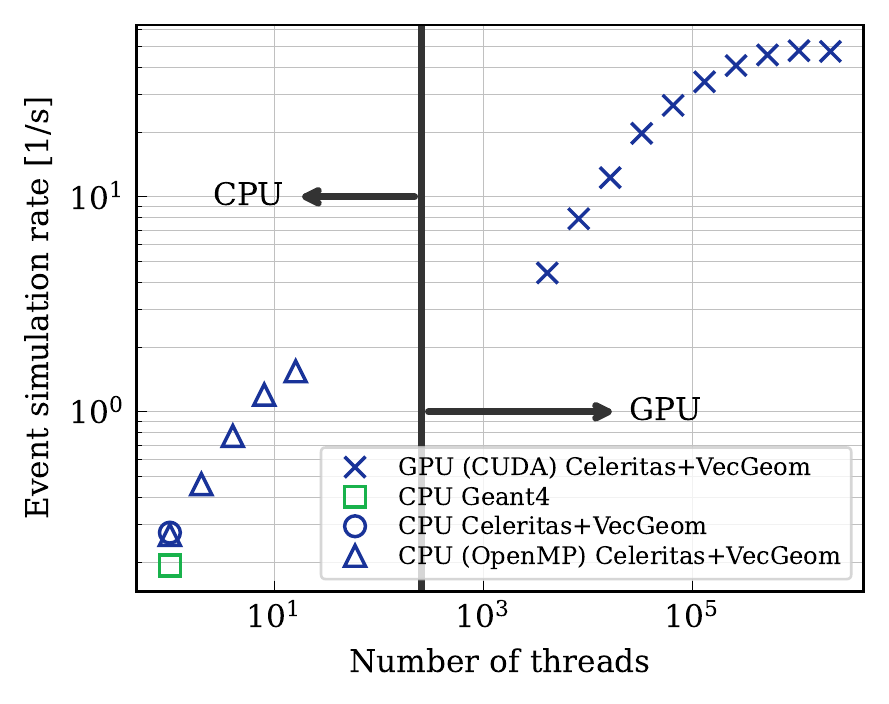}
    \caption{Performance of Celeritas on \acs{cpu} and \acs{gpu} with
  single-core Geant4 performance for reference.}
    \label{fig:celeritas-performance}
\end{figure}

The initial physics validation tallies the cumulative energy deposition over the
$z$ axis (Fig.~\ref{fig:physics-results:edep}), and the distribution of number
of steps per track, separated by particle type
(Fig.~\ref{fig:physics-results:steps}). These show the two \ac{mc} codes to be
in agreement within statistical uncertainties.
The performance numbers can only provide a baseline prediction for the eventual
performance of \celeritas, as it has only a small subset of its eventual
capabilities. Output routines are disabled in order to control for the potential
latency of disk \ac{io}, which will be improved in the future. Because the
current \celeritas application uses the same transport loop algorithm for both
\ac{cpu} and \ac{gpu} implementations, where each step is divided into numerous
kernels with many synchronization points, many tracks must be ``in flight''
simultaneously to amortize the overhead of thread synchronization on \ac{cpu}.
On \ac{gpu}, one thread corresponds to a single track, but the \ac{cpu} can
transport an arbitrary number of tracks per physical core. This problem uses
\num{1024} total tracks regardless of core count on \ac{cpu}. For the results
presented in Fig.~\ref{fig:celeritas-performance}, only one \ac{cpu} socket (up
to 16 physical cores) and a single \ac{gpu} are used. Peak \ac{gpu} performance
is reached with about a million simultaneous tracks. At this peak, the \ac{gpu}
was about $30 \times$ faster than a 16-core run of the OpenMP version of
\celeritas, the equivalent performance of $170$ single-core \ac{cpu} \celeritas
runs or $240$ \acp{cpu} running Geant4.

These initial results are surprisingly good for an initial
unoptimized simulation of a problem with multiple realistic physics,
but they do come with caveats:
\begin{itemize}[itemsep=0pt]
  \item This demonstration version of \celeritas supports only a single element
    per material, and the demonstration problem is hardwired to have no magnetic
    field. Including those features will slow down the simulation even if they
    are unused.
  \item The \ac{prng} used in the \celeritas demonstration ({XORWOW}) is faster
    and less statistically random than the \ac{prng} in Geant4 ({MixMax}).
  \item The current \celeritas algorithm and data structures are designed for
    massive \ac{gpu} parallelism and are not optimal for \ac{cpu} parallelism,
    especially with a small thread count.
  \item Aside from early performance analysis on a small subset of the present
    code base in \cite{johnson_2021}, \emph{no optimization work has been
    performed} for this demonstration, so \celeritas may become faster.
  \item The simulation results are reproducible with the same number of threads,
    but individual track IDs (used for ``MC truth'' debugging analysis) are
    arbitrary.
  \item As with all other \ac{gpu} applications, good performance requires
    saturating the \ac{gpu} with work to do, hiding latency for memory accesses
    and amortizing the launch cost and \ac{cpu} overhead. Experimental workflows
    may need to batch multiple events together to achieve peak \ac{gpu}
    performance.
\end{itemize}
Taking these considerations together, we extrapolate these initial results to be
representative of expected performance on a fully featured \celeritas \ac{em}
simulation.

\begin{figure}
    \centering%
    \begin{subfigure}{\textwidth/2}%
      \includegraphics[height=2.6in]{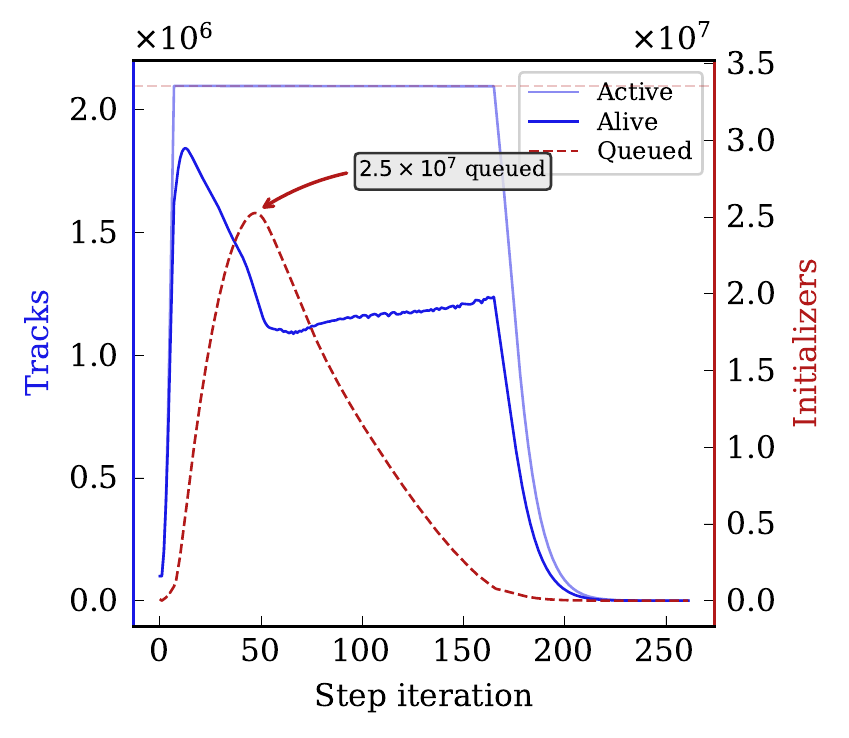}
      \caption{Track and secondaries}%
      \label{fig:gpu-thread-active}%
    \end{subfigure}%
    \begin{subfigure}{\textwidth/2}%
      \includegraphics[height=2.5in]{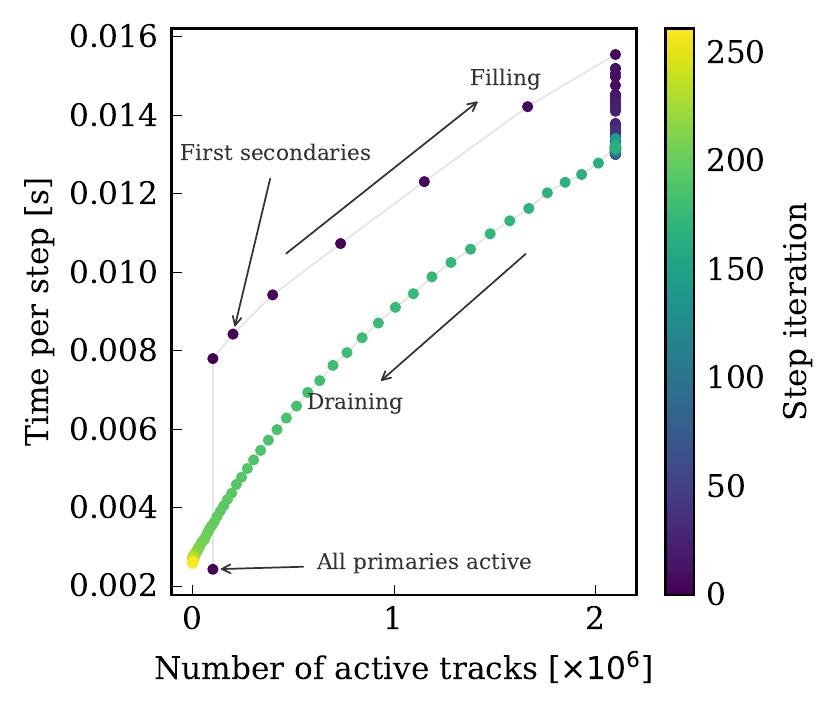}%
      \caption{Performance per step}%
      \label{fig:gpu-thread-time}%
    \end{subfigure}%
    \caption{Characterization on \acs{gpu} of (a) number of tracks in-flight and
    queued, and (b) time taken for each parallel step.}%
    \label{fig:celeritas-steps}%
\end{figure}

Figure~\ref{fig:celeritas-steps} dives into the performance characteristics of
the many-threaded \ac{gpu} execution by enabling extra diagnostics and detailed
timing.
As plotted in Fig.~\ref{fig:gpu-thread-active}, the behavior of the tracks in
flight (blue lines, left axis) and queued secondary initializers (red line,
right axis) determines the maximum thread count in the current implementation,
which is limited by the memory requirements of the initializers: the current
peak of \num{2.5e7} secondaries must at present be below a user-selected
preallocated secondary capacity. For the first ${\sim}10$ steps the total
particle count rises precipitously; but a lower active track count will tend to
complete the transport of low-energy secondaries before starting to transport
the high-energy particles that will increase the total number of tracks in
flight. By preferentially selecting lower-energy particles for tracking,
algorithmic improvements could reduce the total capacity requirement. Additional
code improvements should allow the secondary capacity to reallocate dynamically
and even buffer secondaries in \ac{cpu} memory or \ac{nvm}, maximizing the
number of active tracks and taking full advantage of massive \ac{gpu}
parallelism.

The timing breakdown in Fig.~\ref{fig:gpu-thread-time} zooms in on time
requirements of the steep rise and slow fall of active tracks. The simulation
required seven step iterations to saturate the \ac{gpu} with $2^{21}$ (about two
million) active tracks from the \num{100000} initial primaries. The increase in
time from the initial step (all tracks encounter a boundary since they are born
in a vacuum) to the next step (many tracks interact in matter) is the cost of
running nontrivial physics interaction kernels. Almost a third of the total
iterations are to ``drain'' the pipeline of secondaries. Notably, at the end of
the transport loop, the minimum step iteration time is still more than 10\% of
the maximum step time even with a handful of active tracks. This reiterates the
importance of exposing parallelism for \ac{gpu} performance.

With these initial results we have established a rough projection of the code's
eventual performance and have begun to characterize its behavior on device as a
function of workload. We will continue to use performance results to inform both
low-level code optimizations (to efficiently simulate many tracks in parallel)
and high-level workflow integration (to ensure enough tracks are being simulated
simultaneously).

\section{Integration with \ac{hep} workflows}

\begin{figure}[h]
    \centering
    \includegraphics[width=\textwidth]{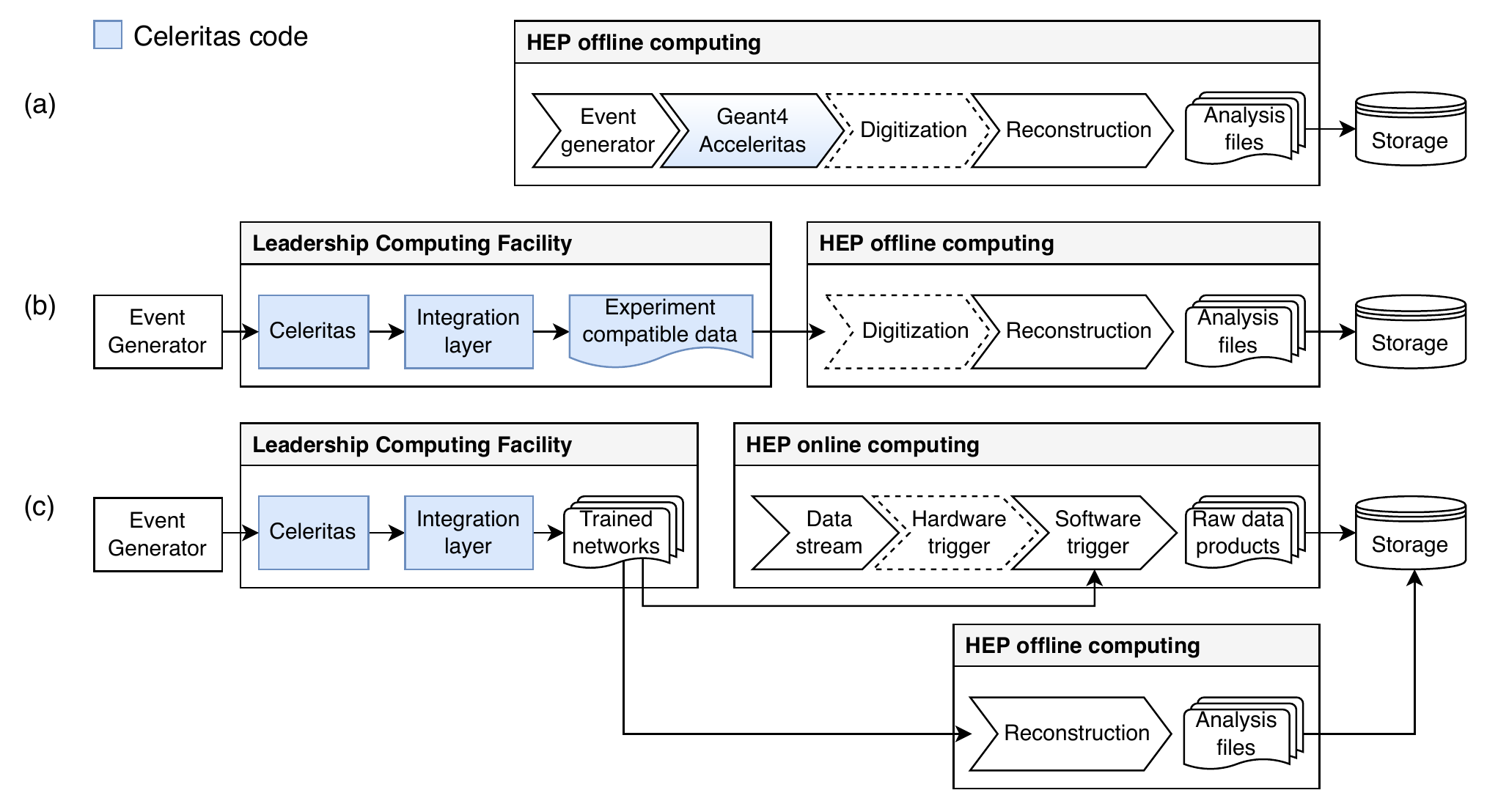}
    \caption{Proposed \acs{hep} integration workflows for (a) \acceleritas, (b)
    end-to-end \celeritas, and (c) \celeritas for \acs{ai}.}
    \label{fig:celeritas-hep-workflows}
\end{figure}

\ac{doe} \acp{lcf} are planned to be part of \ac{hep} workflows by the
scientific community, with their use ranging from simulation and reconstruction
to \ac{ai} methods \cite{hep-network-requirements}. While the Cosmic Frontier is
already taking advantage of facilities such as \ac{alcf}, \ac{nersc}, and
\ac{olcf}, the Energy and Intensity Frontiers have less clear integration
pathways. \celeritas aims close the gap between \ac{hep} distributed computing
networks and \acp{lcf} networks by providing three different routes
(Fig.~\ref{fig:celeritas-hep-workflows}).
These workflows intend to enable \ac{hep} experiments to use \celeritas in three
different ways:
\begin{enumerate}[itemsep=0pt, label=(\alph*)]
  \item \emph{Acceleritas}: accelerate standard \ac{hep} detector simulation
    workflows built on Geant4 by offloading \ac{em} particle showers to
    \acp{gpu} using a new \acceleritas library.
  \item \emph{End-to-end}: run complete end-to-end detector simulations with
    comprehensive \ac{sm} physics at the \acp{lcf}.
  \item \emph{\ac{ai}}: generate high-resolution detector responses as training
    data for \ac{ai} networks to be deployed at experimental facilities as
    software triggers and \ac{ai}-based reconstruction and event selection
    methods.
\end{enumerate}
The next sections will discuss expected challenges and mitigation strategies,
and describe these three envisioned plans to integrate \celeritas into
experimental workflows. Since each workflow has unique traits, we expect that
some experimental collaborations might see value in using all three workflows,
each serving a different purpose.

\subsection{Integration challenges}

The heterogeneity of \ac{hep} computing workflows, associated with the volume of
data produced by each experiment, pose a long list of challenges that need to be
overcome in order to make \celeritas a viable option. We outline here the most
pressing ones, along with mitigation plans.
\begin{enumerate}[itemsep=0pt]
  \item Simulation inputs must be able to encompass Energy and Intensity
    Frontier needs. This includes user-defined geometry, physics, events,
    secondary particle cutoff thresholds, and sensitive detector scoring
    regions.
  \item Output data, which entails \ac{mc} particle history and detector
    scoring, should be flexible enough to make it compatible to experimental
    workflows, while maximizing \ac{io} efficiency.
  \item End-user interface must be simple enough such that the performance gain
    and the work needed to adapt experimental computing workflows justify the
    adoption.
  \item Domestic and international networking between \acp{lcf} and \ac{hep}
    computing centers can lead to large data migration which can result in
    network congestions and suboptimal resource usage.  The \ac{lhc}'s ``any
    data, anywhere, anytime'' model \cite{hep-network-requirements} might need
    special attention.
\end{enumerate}

These main challenges, among other topics, will be discussed and addressed via a
\celeritas \emph{User Council}, which will be formed by members of different
\ac{hep} experiments on both Energy and Intensity Frontiers. The incorporation
\celeritas workflows into existing experimental simulation frameworks will
require early engagement with the experiments. Thus, interactions with the
\emph{User Council} will determine the tradeoffs of applying \celeritas and
\acp{lcf} to their \ac{mc} production, as well as advise the \celeritas team
when developing end-user interfaces, such that the code develops focusing on
experimental compatibility.

\subsection{Acceleritas}
\label{sec:acceleritas}

\acceleritas (Fig.~\ref{fig:celeritas-hep-workflows}a) is a library that will
provide a hybrid workflow between \celeritas and Geant4. It leverages the Geant4
tasking manager system to transfer parts of the simulation to \celeritas for
concurrent execution on device. That will allow a Geant4-based \ac{hep} detector
simulation to collect a subset of particles from either primary collisions or
subsequent hadronic interactions and transport them in parallel on the device
using \celeritas while processing the remainder on the host using Geant4.  The
initial group of offloaded particles will be photons, electrons, and positrons,
but selected hadronic physics will be conditionally integrated based on the
performance of \acceleritas.

The Geant4 tasking manager is responsible for handling all of the major steps in
the process, which includes collecting the list of particles to be offloaded,
launching the \celeritas on-device transport loop, and merging sensitive hits
and particle track data from the device back to Geant4 on the host.

Using Amdahl's law, we expect that the maximum gain of an \acceleritas
application is $1/(1-f)$, where $f$ is the fraction of offloaded work in a
\ac{cpu}-only calculation.  For typical \ac{hep} events at \acs{hllhc} and with
the \ac{cms} detector geometry, the maximum speedup will be roughly a factor 3
for offloading photons and electrons to \acp{gpu} as their fractional \ac{cpu}
contribution is around 70\%.

These gains are nowhere near our expected goals for a full end-to-end \celeritas
simulation, where we expect \celeritas to achieve a similar 160$\times$ speedup
factor observed in Shift. Nevertheless, \acceleritas will provide significant
improvements while requiring minimal adaptations for current \ac{hep} workflows.

\subsection{End-to-end \celeritas}
\label{sec:end-to-end}

\emph{End-to-end} \celeritas integration
(Fig.~\ref{fig:celeritas-hep-workflows}b) requires a mature \celeritas code,
which will rely on the implementation of comprehensive \ac{sm} physics
capabilities along with a fully operational \ac{io} system, including the
possibility to run digitization still at the \acp{lcf}. This incorporation of
\celeritas workflows into existing experimental simulation frameworks will
require early engagement with experiments, and thus depends on a successful
creation and interaction between the \celeritas team with members of the
\emph{User Council}.

An experimental workflow characteristic that will have to be assessed is related
to how event-processing systems are handled on \ac{cpu} and \ac{gpu}. The most
efficient interface would fully occupy the device by executing many events
concurrently.  However, this is not the way that experimental workflows are
currently configured, in which events are executed independently on each thread.
A more seamless integration of the end-to-end workflow would be to preserve
independent event execution.  However, this will have a dramatic effect on the
achievable performance because the \ac{gpu} will have to accumulate sufficient
tracks to fully occupy the device.

The tradeoffs between performance and integration cost will be discussed and
ultimately decided by the experimental collaborations. Finally, the volume of
data transferred between \acp{lcf} and \ac{hep} computing networks is a concern
that will require attention as the project evolves.

\subsection{\celeritas for \ac{ai}}
\label{sec:celeritas-ai}

\ac{ai} methods are now essential in experimental \ac{hep}, with efforts to
unify standard \ac{ai} frameworks with \ac{hep} workflows \cite{mlaas4hep}. The
use of \ac{ai} has been successfully deployed in detector triggering
\cite{ml-trigger}, training surrogate models \cite{fastcalogan}, hit and
image-based reconstruction algorithms \cite{gnn-reco-cms,jets-deep-learning},
and selection of candidate events in data analyses \cite{cvn-nova}. All of these
applications require extensive generation of data for training new networks
before their deployment and, as these techniques are very data and process
intensive, their training can significantly impact \ac{hep} distributed
computing network resources.

With increasing use of \ac{ai} within \ac{hep}, we envision \celeritas and
\acp{lcf} as tools to produce fast, full-fidelity \ac{mc} samples for the
purpose of training new \ac{ai} networks and alleviate the workload on \ac{hep}
computing centers. Standard \ac{ai} frameworks already take advantage of
accelerated architectures, making \celeritas an ideal tool for using \acp{lcf}
as processing centers for \ac{ai} applications in \ac{hep}.

\subsection{The impact of using \acp{lcf} in \ac{hep}}

The current landscape in \ac{hep} computing is evolving. Although the majority
of computing resources will be \ac{cpu}-based for the immediate future,
\ac{gpu}-accelerated hardware will be increasingly added with the advent of
\ac{ai} deep-learning networks in \ac{hep} data analysis and processing
\cite{radovic_machine_2018}. Thus, having \ac{mc} capabilities that uses both
multicore \acp{cpu} and accelerated hardware is a necessity for maximizing use
of available computing resources. Combining current \ac{hep} distributed
computing networks with \acp{lcf} will be sufficient to provide experiments the
capability to rely completely on full-fidelity \ac{mc}, instead of depending on
lower-fidelity fast models due to lack of resources.

\begin{figure}[h]
  \centering%
  \includegraphics[width=0.6\textwidth]{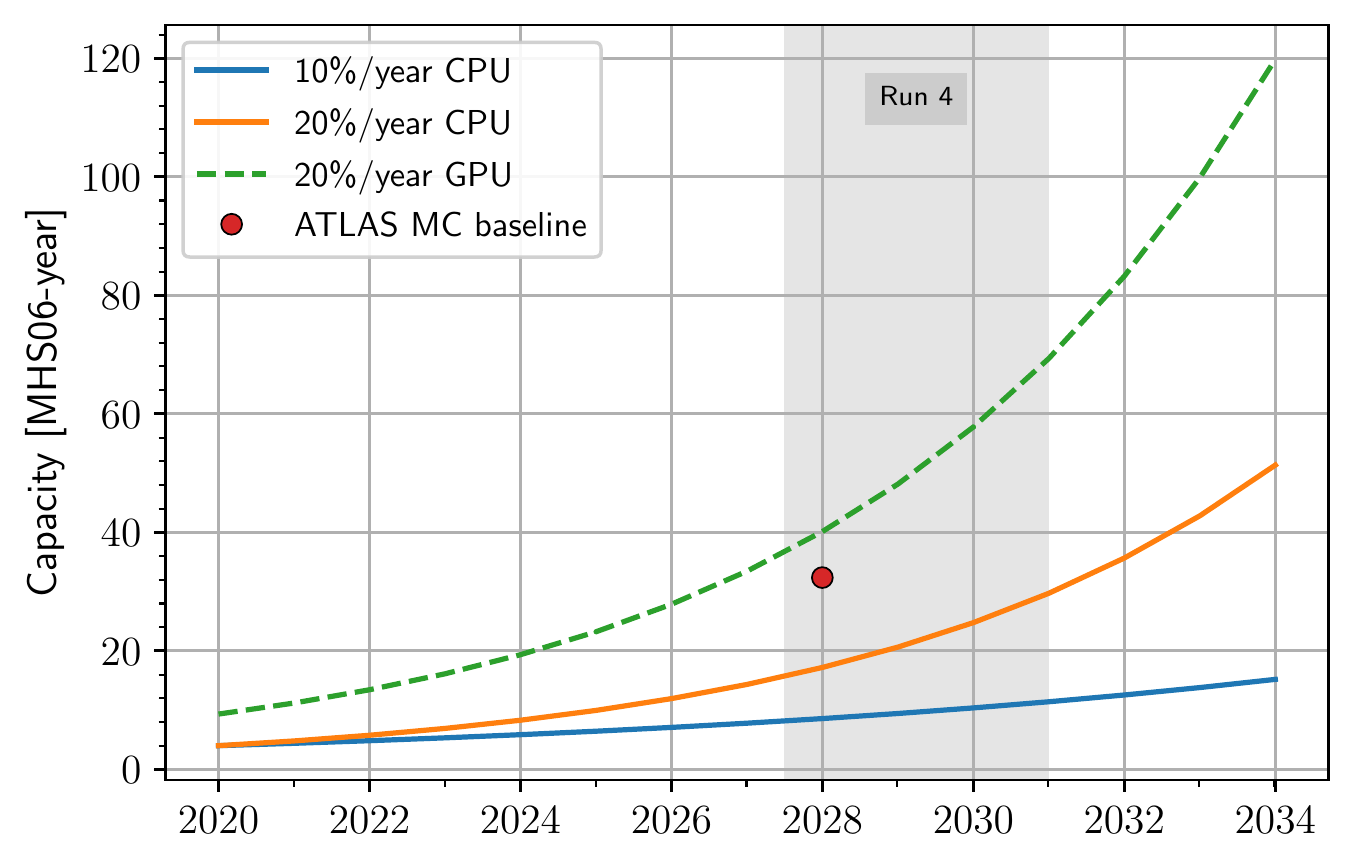}%
  \caption{Compute capability converted into \si{MHS06.year} units for \acs{mc}
  simulation. The 10\% and 20\% \ac{cpu} capacities and 2028 \acs{mc} baseline
  projections are the estimates for \acs{atlas} taken directly from
  Ref.~\cite{the_atlas_collaboration_atlas_2020}.  The green line is the compute
  capacity if all the current computing were converted to \acp{gpu} running at
  equivalent power.}%
  \label{fig:gpu-projection}
\end{figure}

With current solutions, \acs{lhc} experiments are expected to face stringent
computing scenarios by the start of Run 4 \cite{LHC-schedule}. A particular case
is the \acs{atlas} experiment, in which the projected available resources fall
short by a factor of five \cite{the_atlas_collaboration_atlas_2020,
the_hep_software_foundation_roadmap_2019}. To show the impact that \acp{lcf} can
have on experiments if fully integrated, we present a hypothetical projection
comparing the compute capacity of \ac{cpu} vs. \ac{gpu} on \acs{atlas}.
Figure~\ref{fig:gpu-projection} plots the projected \ac{cpu} total resources
(blue and orange lines) and \ac{mc} computing needs for Run 4 (red data point)
using the data provided by Ref.~\cite{the_atlas_collaboration_atlas_2020} and
converts said compute capacity to \ac{gpu} (green dashed line). Here, the
increase in \ac{cpu} capacity is based on 10\% and 20\% hardware annual
improvement assuming a sustained funding, and the \ac{mc} data point represents
39\% of the total computing needs for \acs{atlas}. We use the following
assumptions to project the computing capacity achieved by replacing \acp{cpu}
with \acp{gpu}: \ipl{1} the power consumption from \ac{cpu} computing is
converted into \ac{gpu} capacity; \ipl{2} the compute capability of the
\acp{gpu} will increase 20\% per year through a combination of increased
performance and capacity---this estimated yearly capacity improvement is
justified by recently observed \nvidia performance, which has increased
approximately $1.5\times$ in the last \numrange{3}{4} years; and \ipl{3} each
\ac{gpu} is equivalent to 160 cores using the current performance achieved by
Shift \cite{hamilton_continuous-energy_2019}---a performance metric that we plan
to achieve with \celeritas.

The Summit supercomputer at the \ac{olcf} alone has a total of \num{27648}
\acp{gpu} that, using the aforementioned factor of 160, represents the
equivalent of \num{4423680} \ac{cpu} cores. For comparison purposes, in 2017 the
\ac{wlcg} encompassed approximately \num{500000} \ac{cpu}
cores~\cite{valassi_using_2020}. Incorporating \ac{doe}'s network of \acp{lcf}
will bring the total \ac{hep} computing resources to an unprecedented level and
significantly alleviate current \ac{cpu} bottlenecks.

\section{Conclusion}

We have presented \celeritas, a project designed to provide high-fidelity
\ac{mc} detector simulation transport capabilities on current and next-gen
\ac{gpu} architectures. Over the next years, comprehensive \ac{sm} physics
including \ac{em}, hadronic, and decay physics, along with \ac{em} fields will
be implemented. \celeritas' completion is planned to happen before the beginning
of \acs{hllhc}'s Run 4. This includes a fully fledged \ac{io} system, along with
its integration with ROOT, becoming a viable option to alleviate the impending
\ac{mc} computing requirements of the next generation of \ac{hep} experiments.
With current preliminary results showing performance equivalence between a
single \ac{gpu} and hundreds of \acs{cpu}, \celeritas has the potential to
bring the massive computing power provided by the \ac{doe} \acp{lcf} into
\ac{hep} workflows.

The enabling technologies that will allow interfacing between end-to-end
simulations performed at the \acp{lcf} and experimental computing centers will
yield other long-term benefits for interactions between the \ac{doe} \acp{lcf}
and experimental compute nodes. The planned \ac{io} capabilities that need to be
developed for \celeritas will provide full interoperability between data
produced at the \acp{lcf} and ROOT, which might be benefical for other \ac{hep}
code bases.

The \celeritas project foreshadows proposed efforts in federated computing, in
which the \acp{lcf} interact directly with compute nodes at experimental
facilities to provide optimal use of compute resources.  In this case, we
envision a scenario where expensive \ac{mc} simulations are performed at the
\acp{lcf} and simulation output is communicated directly to experimental
\ac{hep} distributed computing networks for reconstruction and analysis.  The
successful execution of this project can therefore be the genesis for a host of
technological advancements in the use of \ac{hpc} to enable more science output
in all three \ac{doe} \ac{hep} frontiers.

\Acknowledgements Work for this paper was supported by Oak Ridge National
Laboratory (ORNL), which is managed and operated by UT-Battelle, LLC, for the
U.S. Department of Energy (DOE) under Contract No. DEAC05-00OR22725 and by Fermi
National Accelerator Laboratory, managed and operated by Fermi Research
Alliance, LLC under Contract No. DE-AC02-07CH11359 with the U.S. Department of
Energy (Fermilab publication number for this paper is FERMILAB-FN-1159-SCD).
This research was supported by the Exascale Computing Project (ECP), project
number 17-SC-20-SC. The ECP is a collaborative effort of two DOE organizations,
the Office of Science and the National Nuclear Security Administration, that are
responsible for the planning and preparation of a capable exascale
ecosystem---including software, applications, hardware, advanced system
engineering, and early testbed platforms---to support the nation's exascale
computing imperative.

\printbibliography

\end{document}